\documentclass[twoside]{article}
\usepackage{qic,epsf}

\pdfoutput=1
\usepackage{amsmath}
\usepackage{graphicx}
\usepackage{txfonts}
\usepackage{amsfonts}
\usepackage{amssymb}
\usepackage{graphicx}%
\usepackage{color}
\usepackage{tikz}%
\usetikzlibrary{decorations.pathreplacing}
\usetikzlibrary{fit}
\providecommand{\U}[1]{\protect\rule{.1in}{.1in}}
\def\H{{\mathcal H}}

\begin{document}
\setlength{\textheight}{8.0truein}    %FOR 2ND PAGE ONWARDS
\copyrightheading{12}{5\&6}{2012}{0394-0402}
\runninghead{Encryption with Weakly Random Keys Using Quantum
Ciphertext  }
            {J. Bouda, M. Pivoluska, M. Plesch }

\normalsize\textlineskip
\thispagestyle{empty}
\setcounter{page}{1}

\vspace*{0.88truein}

\alphfootnote

\fpage{1}

\centerline{\bf
%%%%%%%%%%%%%%%%%%%%%
%Put in titles here
%%%%%%%%%%%%%%%%%%%%%
ENCRYPTION WITH WEAKLY RANDOM KEYS USING A QUANTUM CIPHERTEXT}
\vspace*{0.035truein}

\vspace*{0.37truein}
\centerline{\footnotesize
%%%%%%%%%%%%%%%%%%%%%%%%%%%%%%%%%%%%
%put authors' name and address here
%%%%%%%%%%%%%%%%%%%%%%%%%%%%%%%%%%%%
JAN BOUDA}
\vspace*{0.015truein}
\centerline{\footnotesize\it Faculty of Informatics, Masaryk University, Botanick\'{a} 68a}
\baselineskip=10pt
\centerline{\footnotesize\it Brno, Czech republic}
\vspace*{10pt}
\centerline{\footnotesize
MATEJ PIVOLUSKA}
\vspace*{0.015truein}
\centerline{\footnotesize\it Faculty of Informatics, Masaryk University, Botanick\'{a} 68a}
\baselineskip=10pt
\centerline{\footnotesize\it Brno, Czech republic}
\vspace*{10pt}
\centerline{\footnotesize
MARTIN PLESCH}
\vspace*{0.015truein}
\centerline{\footnotesize\it Faculty of Informatics, Masaryk University, Botanick\'{a} 68a}
\baselineskip=10pt
\centerline{\footnotesize\it Brno, Czech republic}

\centerline{\footnotesize\it Institute of Physics, Slovak Academy
of Sciences, D\'{u}bravsk\'{a} cesta 9} \baselineskip=10pt
\centerline{\footnotesize\it Bratislava, Slovakia} \vspace*{10pt}
\vspace*{0.225truein} \publisher{(June 17, 2011)}{(January 19,
2012)}

\vspace*{0.21truein}

%% \abstracts{first paragraph}{second paragraph}{third paragraph}
%% If there is only one paragraph, just keep the second and third empty
%% like the following one

\abstracts{ The lack of perfect randomness can cause
significant problems in securing communication between two
parties. McInnes and Pinkas
\cite{McInnesPinkas-ImpossibilityofPrivate-1991} proved
that unconditionally secure encryption is impossible when
the key is sampled from a weak random source. The adversary
can always gain some information about the plaintext,
regardless of the cryptosystem design. Most notably, the
adversary can obtain full information about the plaintext
if he has access to just two bits of information about the
source (irrespective on length of the key). In this paper
we show that for every weak random source there is a
cryptosystem with a classical plaintext, a classical key,
and a quantum ciphertext that bounds the adversary's
probability $p$ to guess correctly the plaintext strictly
under the McInnes-Pinkas bound, except for a single case,
where it coincides with the bound. In addition, regardless
of the source of randomness, the adversary's probability
$p$ is strictly smaller than $1$ as long as there is some
uncertainty in the key (Shannon/min-entropy is non-zero).
These results are another demonstration that quantum
information processing can solve cryptographic tasks with
strictly higher security than classical information
processing.} {}{}
%\pacs{TB never A}

%%%%%%%%%%%%%%%%%%%%%%%%%%%%%%%%%%%%%%%%%%%%%%%%%%%%%%%%%%%%%%%%%%%%%%%%%%%%%%%

\section{Introduction}

%%%%%%%%%%%%%%%%%%%%%%%%%%%%%%%%%%%%%%%%%%%%%%%%%%%%%%%%%%%%%%%%%%%%%%%%%%%%%%%

Random numbers play a crucial role in many areas of
computer science, e.g. randomized algorithms and
cryptography. Real world random number generators deliver
imperfect (biased) randomness and a number of theoretical
models of imperfect random number sources
\cite{Blum-Independentunbiasedcoin-1984,ChorGoldreich-Unbiasedbitsfrom-1988,SanthaVazirani-Generatingquasi-randomsequences-1985,Neumann-Varioustechniquesused-1951}
were introduced to study possibilities to obtain perfect
randomness through their software postprocessing
\cite{Shaltiel-RecentDevelopmentsin-2002}. Importance of
this post-processing stems from the fact that many of
its applications were designed to use perfect randomness, and,
in fact, vitally require it for a reasonable performance.
Devices based on quantum mechanical properties should
theoretically serve as sources of ideal randomness
\cite{ColbeckRenner-2011,idQuantique1} or even as unconditionally secure
cryptosystems \cite{BennetBrassard-1984,idQuantique2,Mayers-2001,ShorPreskill-2000,TamakiKoashiImoto-2003}. However, in real
conditions they are extremely sensitive to the influence of
environment and rely on classical post-processing of the
measurement results. Even after these procedures the
outcome is far from being perfect (see e.g.
\cite{Lydersen-2010} and references therein).

Cryptography counts among fields that are highly sensitive
to the quality of randomness used (see. e.g.
\cite{Dodis-2002}). One of the most prominent results
showing the influence of weakness of randomness is by
McInnes and Pinkas
\cite{McInnesPinkas-ImpossibilityofPrivate-1991} proving
that there is no perfectly secure encryption scheme when
the key is sampled from a weakly random source (e.g.
min-entropy \cite{ChorGoldreich-Unbiasedbitsfrom-1988}).
The authors also derived a tight (minimal and achievable) bound
on probability that the adversary can determine the
plaintext, when an arbitrary encryption system is used, but
the key is sampled from a min-entropy source. This bound is
a function of $c = l-b$, where $l$ is the key length and
$b$ is its min-entropy (see Section 2 for definition of
min-entropy).

In this paper we propose an encryption of classical
information using classical key and quantum channel (i.e.
ciphertext is a quantum state) such that the adversary's
probability to determine the plaintext is strictly smaller
than the McInnes-Pinkas bound for all values $c=l-b\geq0$,
except for values $c=0$ and $c=1$, where it coincides with
the bound.

The paper is organized as follows. In the second section we
introduce basic definitions and recall the result by
McInnes and Pinkas in detail. In the third section we introduce
the encryption onto a quantum ciphertext. In the fourth
section we derive the maximal security for the
approximation of a continuous key. In the fifth section we
deal with the consequences of the discretisation of the
random key, whereas in the last section we summarize our
results and conclude.

%%%%%%%%%%%%%%%%%%%%%%%%%%%%%%%%%%%%%%%%%%%%%%%%%%%%%%%%%%%%%%%%%%%%%%%%%%%%%%%

\section{Preliminaries}

%%%%%%%%%%%%%%%%%%%%%%%%%%%%%%%%%%%%%%%%%%%%%%%%%%%%%%%%%%%%%%%%%%%%%%%%%%%%%%%

The min-entropy of the probability distribution (random variable, source)
$\mathbf{Z}$ is defined by
\begin{equation}\label{minentropy}
H_{\infty}(\mathbf{Z})=\min_{z\in Z}\left(  -\log Pr\left(  \mathbf{Z}%
=z\right)  \right)  .
\end{equation}
We denote a source as $(l,b)$-source if it is emitting $l$-bit strings drawn
according to a probability distribution with min-entropy at least $b$. Thus,
every specific $l$-bit sequence is drawn with probability smaller or equal to
$2^{-b}$. Notice that for $b=l-c$, the probability of each $l$-bit string is upper bounded by $2^c\frac{1}{2^{-l}}$.
and parameter $c$ is called min-entropy loss.

A source is $(l,b)$-flat iff it is an $(l,b)$ source and it is uniform on some
subset of $2^{b}$ sample points i.e., all probabilities
are either $0$ or $2^{-b}$.

We are going to consider the following scenario. Alice and
Bob share a secret key $k$ that is used to determine the
encoding (decoding) function. In this setting the plaintext
is a single, uniformly distributed bit and the ciphertext
is (arbitrarily long, but finite) bitstring. The encryption
system is specified by the set of encoding rules
\begin{equation}
e_{k}:X\rightarrow Y
\end{equation}
parameterized by the keys $k\in K$, with $X=\{0,1\}$ being the set of
plaintexts and $Y$ being the set of ciphertexts. To each encoding function
there is the corresponding decoding function $d_{k}$ that perfectly recovers
the original input, i.e.
\begin{equation}
\forall k\in K\ d_{k}\circ e_{k}=id_{X}.
\end{equation}

Let ${\mathbf{X}}$ be the random variable describing the probability
distribution of the plaintext and ${\mathbf{X}^{\prime}}$ random variable
describing adversary's estimate given by his decoding strategy. Security is
parameterized by adversary's ability to recover the original plaintext
\begin{equation}
\label{equ:adversary_probability}p=\sum_{x\in X}Pr({\mathbf{X}^{\prime}%
}=x|{\mathbf{X}}=x)Pr({\mathbf{X}}=x).
\end{equation}

In the McInnes and Pinkas paper, for a given length of the
key $l$ and a parameter $c$ the key is distributed
according to an $(l,l-c)$ distribution. Parameters $l$ and
$c$ are part of the cryptosystem design, the adversary is
assumed to know the actual (biased) distribution
${\mathbf{K}}$ of the key (but not the value $k$ of the
key). The probability $p$ to reveal the plaintext the
adversary can achieve for an arbitrary cryptosystem, a suitable
$(l,l-c)$ source and a uniformly distributed plaintext is
lower bounded by
\begin{equation}
p\geq%
\begin{cases}
1 & \text{for }2\leq c\leq l\\
\frac{1}{2}+\frac{2^{c}}{8} & \text{for }2-\log_{2}3\leq c\leq2\\
\frac{2^{c}}{2} & \text{for }0\leq c\leq2-\log_{2}3.
\end{cases}
\label{equ:pinkas_probability}%
\end{equation}

In particular, they have shown that the (maximal achievable)
security\footnote{Probability of making a correct guess by the adversary is
bounded from below by the formula \eqref{equ:pinkas_probability} independently
of $l$. However, such security is achievable only for a cryptosystem with high
enough $l$, for smaller $l$ the achievable probability for the attacker rises
further.} is independent of $l$ and no security can be achieved if $c\geq2$.

%%%%%%%%%%%%%%%%%%%%%%%%%%%%%%%%%%%%%%%%%%%%%%%%%%%%%%%%%%%%%%%%%%%%%%%%%%%%%%%

\section{Quantum ciphertext}

%%%%%%%%%%%%%%%%%%%%%%%%%%%%%%%%%%%%%%%%%%%%%%%%%%%%%%%%%%%%%%%%%%%%%%%%%%%%%%%

In our solution we consider encoding classical plaintext to a quantum
ciphertext. The encoding function is of the form
\begin{equation}
e_{k}:X\rightarrow{\mathrm{S}}({\mathcal{H}})
\end{equation}
parameterized by the key $k\in K$, with $X=\{0,1\}$ being the set of
plaintexts, ${\mathcal{H}}$ a suitable Hilbert space and ${\mathrm{S}%
}({\mathcal{H}})$ the set of (possibly mixed) states on the
Hilbert space ${\mathcal{H}}$, i.e. positive trace one
operators acting on ${\mathcal{H}}$. In our further
analysis we limit ourselves to only a single quantum bit,
i.e. to two-dimensional Hilbert space
${\mathcal{H=H}}_{2}$. Extensions to higher dimensional
Hilbert spaces will be discussed in the conclusion.

The decoding procedure consists of measurement of the
received quantum system aiming to distinguish between
states $e_{k}(0)$ and $e_{k}(1)$. The correctness
requirement gives (regardless of ${\mathcal{H}}$) that for
every $k$ the states $e_{k}(0)$ and $e_{k}(1)$ must be
orthogonal. For a qubit, this is only possible if all of
them are pure states. Let us use the notation
$e_{k}(0)=\left\vert \psi_{k}\right\rangle $ and
$e_{k}(1)=\left\vert \phi
_{k}\right\rangle $, with the orthogonality condition%
\begin{equation}
\left\langle \phi_{k}|\psi_{k}\right\rangle =0\label{orthogonality}%
\end{equation}
for all $k$'s.

To obtain the encoded bit, Bob (knowing the key) adjusts his measurement
device accordingly to obtain a well defined result discriminating between
$\left\vert \psi_{k}\right\rangle $ and $\left\vert \phi_{k}\right\rangle $.
For this purpose a standard von Neumann measurement is fully sufficient. On
the other hand, the adversary's knowledge is limited to the (known) probability
distribution on keys. He has to discriminate between the average states
\begin{equation}
\rho_{0}=\sum_{k\in K}P({\mathbf{K}}=k)\left\vert \psi_{k}\right\rangle\left\langle\psi_k\right\vert
\text{ and }\rho_{1}=\sum_{k\in K}P({\mathbf{K}}=k)\left\vert \phi_{k}\right\rangle\left\langle\phi_k\right\vert .
\label{equ:average states}%
\end{equation}
Analogously to the classical case, the effectiveness of the
adversary's strategy is given by the probability $p$ given by Eq.
\eqref{equ:adversary_probability}. The adversary's strategy is
thus to maximize this probability by performing a minimum error
measurement, which is a simple two outcome von Neumann measurement
\cite{Heinosaari-2009,Helstrom-1976,Nielsen+Chuang-Quant_Compu_Quant:2000}.

Without the loss of generality we can choose a basis so that the state
$\rho_{0}$ (\ref{equ:average states}) has the form
\begin{equation}
\rho_{0}=\left(
\begin{array}
[c]{cc}%
a & 0\\
0 & 1-a
\end{array}
\right)
\end{equation}
with $a\geq\frac{1}{2}$. Due to the orthogonality condition
(\ref{orthogonality}) the state $\rho_{1}$ has the form
\begin{equation}
\rho_{1}=\left(
\begin{array}
[c]{cc}%
1-a & 0\\
0 & \ a\
\end{array}
\right)  .
\end{equation}
The optimal measurement of the attacker is thus just the spin $z$ projection
measurement and the probability to get a correct outcome (determine the
original plaintext) then reads%
\begin{equation}
p=\frac{1}{2}\left[  \frac{1}{2}Tr\left\vert \rho_{0}-\rho_{1}\right\vert
+1\right]  =a.\label{adversary_probability_quantum}%
\end{equation}

%%%%%%%%%%%%%%%%%%%%%%%%%%%%%%%%%%%%%%%%%%%%%%%%%%%%%%%%%%%%%%%%%%%%%%%%%%%%%%%

\section{Continuous code}
%%%%%%%%%%%%%%%%%%%%%%%%%%%%%%%%%%%%%%%%%%%%%%%%%%%%%%%%%%%%%%%%%%%%%%%%%%%%%%%

In this section we assume that Alice and Bob share a random
key that is continuous, e.g. a single complex number. Such
a coding can uniformly cover the state space of a single
qubit $\mathcal{H}$, which can be depicted as a Bloch
sphere. Later, in Section \ref{sec:discrete_code}, we will
show that there exists a discrete coding that approximates
the continuous coding introduced in this section with an
arbitrary precision (with respect to adversary's
probability to reveal the plaintext).

Suppose that Alice and Bob know (or expect) that the random
key they share can be biased by a certain amount. Their aim
is to choose the coding in such a way that any partial
knowledge about the key by an eavesdropper would lead to as
small probability of obtaining the correct encrypted bit as
possible. This is naturally achieved by a smooth coverage
of the state space (Bloch sphere) in such a way that the
probability density of selected states will be equal on all
points of the sphere.

The first important observation is that we can fix an arbitrary
adversary's measurement $P={|0\rangle}{\langle0|}$ by fixing the
basis and determine the key distribution that is optimally
distinguished by the measurement. This can be done as all
measurements are unitarily equivalent, i.e. for each pair of
measurements there is a unitary rotation of the sphere that maps
one measurement to the other. Hence, if there is an optimal
distribution for one measurement, for any other measurement there
exists a (different) distribution giving the same result.

Let us define a source with min-entropy loss at most $c$ for
continuous spaces. According to \eqref{minentropy} a discrete
source with length $l$ and min-entropy $H_\infty(\mathbf{Z})$ has
min-entropy loss $c=l-\mathbf{H}_\infty(\mathbf{Z})$:
\begin{align}
\nonumber
c& = l - H_\infty\left(\mathbf{Z}\right)\\
\nonumber
& = l - \min_{z\in \mathbf{Z}} \left(-\log Pr\left(\mathbf{Z}=z\right)\right)\\
\nonumber
& = \max_{z\in \mathbf{Z}}  \left(l +\log Pr\left(\mathbf{Z}=z\right)\right)\\
& = \max_{z\in \mathbf{Z}}  \left(\log
\left(|\mathbf{Z}|Pr\left(\mathbf{Z}=z\right)\right)\right).\label{discML}
\end{align}

Equation \eqref{discML} can be easily extended to continuous space
by changing maximalization to $\sup$. $|\mathbf{Z}|$ has to be
changed to the volume of the probability space $\H$ and
probability function $Pr$ becomes probability density function
$\mu$.

Now let us define a continuous weak source over space $\H$ with min entropy loss $c$ as a set of probability density functions
for which
\begin{equation}
c = \sup_{\psi\in\H}\log\left({|\H|\mu\left(\phi\right)}\right).
\end{equation}
After additional simplification it is easy to see that the
condition reads $\sup_{\H}\mu\left(\phi\right)=\frac{2^c}{|\H|}$.

Let us consider all distributions on the sphere such that for any
state on the sphere its probability density is at most
$2^c{|\H|}^{-1}$, with $|\H|$ being the area of the surface of the
Bloch sphere. Later, in Section \ref{sec:discrete_code}, we show
that such distributions are analogous to discrete min-entropy
sources. Flat distributions correspond to continuous distributions
where only a $2^{-c}$ fraction of all possible keys would appear
with equal nonzero probability density.

Let us now examine a situation, where the adversary prepares for Alice and Bob
a flat distribution on the subset of size $2^{-c}\left\vert {\mathcal{H}%
}\right\vert $ (i.e. the keys are selected with equal probabilities from
$2^{-c}$ fraction of all possible keys). We propose a distribution $\mu
_{opt}\left(  {|\phi\rangle}\right)  $ defined as%
\[
\mu_{opt}\left(  {|\phi\rangle}\right)  =%
\begin{cases}
\frac{1}{2^{-c}\left\vert {\mathcal{H}}\right\vert } & \text{for }%
{|\phi\rangle}\in Y\\
0 & \text{elsewhere}%
\end{cases}
\]
for $Y=\{{|\phi\rangle}\in{\mathcal{H}};\left\vert \langle0|\phi
\rangle\right\vert ^{2}\geq g;|Y|=2^{-c}\left\vert
{\mathcal{H}}\right\vert\}$ for some suitable constant $g$
dependent on $c$. Let us assume that Alice encodes $0$ (uniformly
at random) into one of the states from $Y$. The probability to
obtain the correct outcome of the measurement $P$ is
\begin{align}
p_{opt} &  =\int_{{\mathcal{H}}}\mu_{opt}\left(
{|\phi\rangle}\right) \left\vert \langle0|\phi\rangle\right\vert
^{2}d\phi\label{p_opt}   =\frac{1}{2^{-c}\left\vert
{\mathcal{H}}\right\vert }\int_{{|\phi\rangle }\in Y}\left\vert
\langle0|\phi\rangle\right\vert ^{2}d\phi.
\end{align}

Let $\mu\left(  {|\phi\rangle}\right)$ be any distribution with
$\mu\left(  {|\phi\rangle}\right)\leq 2^{c}|\H|^{-1}$ everywhere,
i.e. we restrict ourselves to distributions corresponding to
discrete min-entropy $(l,l-c)$ distributions. We will now prove
that the distribution $\mu_{opt}$ is optimal among these
distributions, i.e there is no other distribution such that the
adversary can obtain a higher probability to detect the correct
ciphertext. Let us calculate this probability for a general
distribution $\mu\left({|\phi\rangle}\right)  $%

\begin{align}
p &  =\int_{{\mathcal{H}}}\mu\left(  {|\phi\rangle}\right)  \left\vert
\langle0|\phi\rangle\right\vert ^{2}d\phi\nonumber\\
&  =\int_{{|\phi\rangle}\in Y}\mu\left(  {|\phi\rangle}\right)  \left\vert
\langle0|\phi\rangle\right\vert ^{2}d\phi+\int_{{|\phi\rangle}\in\backslash
Y}\mu\left(  {|\phi\rangle}\right)  \left\vert \langle0|\phi\rangle\right\vert
^{2}d\phi\nonumber\\
&  =\int_{{|\phi\rangle}\in Y}\mu_{opt}\left(  {|\phi\rangle}\right)  \left\vert
\langle0|\phi\rangle\right\vert ^{2}d\phi+\int_{{|\phi\rangle}\in Y}\left[
\mu\left(  {|\phi\rangle}\right)  -\mu_{opt}\left(  {|\phi\rangle}\right)
\right]  \left\vert \langle0|\phi\rangle\right\vert ^{2}d\phi+\int
_{{|\phi\rangle}\in\backslash Y}\mu\left(  {|\phi\rangle}\right)  \left\vert
\langle0|\phi\rangle\right\vert ^{2}d\phi\nonumber\\
&  =p_{opt}+\int_{{|\phi\rangle}\in Y}\left[  \mu\left(  {|\phi\rangle
}\right)  -\mu_{opt}\left(  {|\phi\rangle}\right)  \right]  \left\vert
\langle0|\phi\rangle\right\vert ^{2}d\phi+\int_{{|\phi\rangle}\in\backslash
Y}\mu\left(  {|\phi\rangle}\right)  \left\vert \langle0|\phi\rangle\right\vert
^{2}d\phi,\label{flat}%
\end{align}
where $\backslash Y=\mathcal{H}-Y$.

Let us now analyze the
two remaining terms in the Eq. (\ref{flat}). As $\mu\left(
{|\phi\rangle}\right) \leq\mu_{opt}\left(
{|\phi\rangle}\right) $ for all ${|\phi\rangle}\in Y$
(recall that $\mu_{opt}\left( {|\phi\rangle}\right)  $
reaches the maximal permitted value on the subset $Y$), the
integrated function is non-positive on the whole area of
integration ($Y$), $\left\vert \langle0|\phi\rangle\right\vert
^{2}\geq g$ on the whole area as well and thus the first integral
can be bounded from above by $g\int
_{{|\phi\rangle}\in Y}\left[ \mu\left(
{|\phi\rangle}\right)  -\mu _{opt}\left(
{|\phi\rangle}\right)  \right]  d\phi$. In the last term the
integrated function is positive, but $\left\vert
\langle0|\phi\rangle \right\vert ^{2}<g$ on the whole area
and thus the second integral can be also bounded from above
by $g\int_{{|\phi\rangle}\in\backslash Y}\mu\left(
{|\phi\rangle}\right)  d\phi$. Altogether we get%
\[
p\leq p_{opt}+g\int_{{|\phi\rangle}\in Y}\left[  \mu\left(  {|\phi\rangle
}\right)  -\mu_{opt}\left(  {|\phi\rangle}\right)  \right]  d\phi
+g\int_{{|\phi\rangle}\in\backslash Y}\mu\left(  {|\phi\rangle}\right)  d\phi.
\]
As $\mu_{opt}\left(  {|\phi\rangle}\right)  $ is $0$
outside $Y$, we can include it into the integration outside
$Y$ with a minus sign. Now we join the
integration through the whole space and using the normalization condition on both distributions we get%
\[
p\leq p_{opt}+g\int_{{\mathcal{H}}}\left[  \mu\left(
{|\phi\rangle}\right) -\mu_{opt}\left(
{|\phi\rangle}\right)  \right]  d\phi=p_{opt}.
\]
This completes the proof of the optimality of the flat
$\mu_{opt}\left(  {|\phi\rangle}\right)  $ distribution.

Recall that the state space of a single qubit represented as a Bloch sphere is
proportional to a unit sphere with the surface equal to $\left\vert
{\mathcal{H}}\right\vert =4\pi$. The set $Y$ derived above is a spherical cap
on the Bloch sphere with the center in ${|0\rangle}$. The desired flat
distribution is hence equivalent to the uniform distribution on a spherical
cap with a surface $4\pi2^{-c}$. The height of such a spherical cap is
$h=2^{-c+1}$ (reaching from $2$ for $c=0$ to $0$ for large $c$). The average
state observed by the attacker in the case of the plaintext $0$ is the center of
the mass of the surface of the spherical cap, which is on the axis of the cap
at the height $h/2$. The average state then reads
\begin{equation}
\rho_{0}=\frac{1}{2}\mathbb{I}+\frac{h}{2}{\sigma}_{z}=p{|0\rangle}{\langle
0|}+\left(  1-p\right)  {|1\rangle}{\langle1|}%
\end{equation}
with
\begin{equation}
p=1-h/4=1-2^{-c-1}.\label{probability_quantum}%
\end{equation}
The appearance of $h/4$ instead of $h/2$ is due to the renormalization of the
axis: while the Bloch sphere has diameter $2$, the parameter $p$ changes from
$0$ to $1$ across the sphere.

Accordingly, the state observed by the attacker in the case of sending $1$ is
$\rho_{1}=\left(  1-p\right)  {|0\rangle}{\langle0|}+p{|1\rangle}{\langle1|}$.
Observing that $p\geq1/2$ and substituting into Eq.
(\ref{adversary_probability_quantum}), the optimal adversary's probability for
this state is also $p$.

\begin{figure}[ptb]
\begin{center}
%l\epsfig{file=figure.eps, width=2.399900in}
\includegraphics[
natheight=1.619800in, natwidth=2.399900in, height=1.619800in,
width=2.399900in]{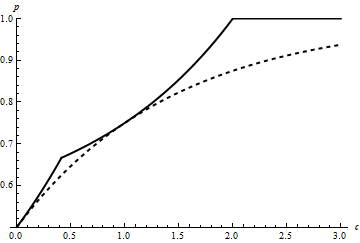}
\end{center}
\fcaption{Comparison of the probability of a successful
guess by the attacker, if Alice and Bob use classical
channel (full line) and quantum channel (dashed
line). }%
\label{figure1}%
\end{figure}

The comparison of classical bounds and quantum approach is given in the Fig.
(\ref{figure1}). It is clear that the probability of the adversary to
correctly guess the ciphertext is for all parameters $c$ (except $0$ and $1$) strictly better
than classical. Even more interesting, the probability is non-vanishing even
for large $c$, what means that (some) privacy is established even in the case
when the adversary has almost perfect control of the "random" key.

%A question may arise whether Alice and Bob could potentially increase their
%privacy by choosing a non-uniform mapping between the keys and the state
%space. The answer is, as expected, negative. In such a case the unitary
%equivalence of measurement would not longer hold and the adversary could
%choose a measurement direction, under which a subset of $2^{-c}\left\vert
%{\mathcal{H}}\right\vert $ states would be concentrated in a spherical cap
%with a height smaller than $2^{-c+1}$. As a consequence, the adversary would be
%able to determine the ciphertext with higher probability.

%%%%%%%%%%%%%%%%%%%%%%%%%%%%%%%%%%%%%%%%%%%%%%%%%%%%%%%%%%%%%%%%%%%%%%%%%%%%%%%

\section{Discrete code}
\label{sec:discrete_code}
%%%%%%%%%%%%%%%%%%%%%%%%%%%%%%%%%%%%%%%%%%%%%%%%%%%%%%%%%%%%%%%%%%%%%%%%%%%%%%%

In this section we will show that for any $c\geq0$ and an
arbitrarily small $\epsilon>0$ there exists an encryption system
\begin{equation}
e_{k}:\{0,1\}\rightarrow{\mathrm{S}}({\mathcal{H}})
\end{equation}
indexed by keys from a finite set $K$ that bounds the adversary's
probability to determine the original plaintext by
$1-2^{-c-1}+\epsilon$.

 All the derivations made in the previous section were done for
continuous distribution of the states $\left\vert
\psi\right\rangle $ over the state space. If we consider a
discrete finite number $n$ of possible keys $\left\vert
\psi_{i}\right\rangle $, we can only approximate such distributions
using a finite number of lattice points
\cite{Kraetzel-LatticePoints-1989} uniformly distributed over the
surface of the Bloch sphere.

 Let us assign to each single state $\left\vert \psi
_{i}\right\rangle $ its neighborhood, i.e. a subset
${\mathcal{H}}_{i}$ of the whole Hilbert space such that

\begin{itemize}
\addtolength{\itemsep}{-2mm}

\item[(i)] The surface on the Bloch sphere of each
${\mathcal{H}}_{i}$ equals to $\frac{4\pi}{n}$.

\item[(ii)]  Neighborhoods corresponding to different states are
disjoint, i.e. $\forall
i,j\,{|\psi_{i}\rangle}\neq{|\psi_{j}\rangle
}\Rightarrow{\mathcal{H}}_{i}\cap{\mathcal{H}}_{j}=\emptyset$.

\item[(iii)]  The distance between a state $\left\vert
\psi_{i}\right\rangle $ and any state in its neighborhood measured
in the relative angle on the Bloch sphere is upper bounded by
$O\left( n^{-1/2}\right)  $. \end{itemize}

For every $n$, there exist a set of states $\left\vert \psi
_{i}\right\rangle $ and its neighborhoods that fulfill the
aforementioned conditions. This can be seen from the fact that
even a suitable (by far non-optimal) distribution of points on
$\sqrt{n}$ meridians with $\sqrt{n}$ points each yields a maximal
angle of $\frac{3\pi}{2\sqrt{n}}$ (for explanation see Fig.
(\ref{poludnik})), which fulfills the third condition, whereas the
first two can be fulfilled trivially.

 \begin{figure}[ptbh]
\centering \label{poludniky}
\begin{tikzpicture}[scale=0.6]
\draw (0,-1) arc (180:360:4 and .6);
\draw[loosely dashed] (8,-1) arc (0:180:4 and .6);

\draw[color=black,<->] (3,0.05) arc (164:184:1 and 4)node[yshift = 0.5cm, xshift = -0.3cm] {$b$};
\draw[color=black,<->] (3.3,-1.6) arc (260:281:4 and 1)node[yshift = -0.3cm, xshift = -0.43cm] {$a$};
\draw (4,3) arc (90:270:1 and 4);
\draw (4,-5) arc (-90:90:1 and 4);
\draw (4,-1) circle (4);
\node[circle, fill=black!30] (n1) at (3,-1.6) {};
\node[circle,fill=black!30] (n2) at (5,-1.6) {};
\node[circle, fill=black!30](n3) at (3.1,0.3)  {};
\node[circle, fill=black!30] (n4) at (4.9,0.3){};
\node[circle, fill = black] (n5) at (4,-0.6)  {};
\draw[<->] (3.1,-1.5) --
(3.8,-0.8)node[black,midway,xshift=0.15cm,yshift = -0.15cm]  {$c$};
\end{tikzpicture}
\fcaption{As there are $\sqrt{n}$ meridians used, there are
$\sqrt{n}$ points on the equator. Naturally the most distant
points are in the vicinity of the equator, as there are the
meridians most distant from each other. On the picture,  four
points are depicted, two of them on meridian and two just one
position above. The most distant point to any of these points is
in the center of the formation formed by these four points. The
distance $c$ is bounded from above by the triangle inequality by
its distance to the equator $\frac{b}{2}=\frac{\pi}{2\sqrt{n}}$
and half of the distance between two points on the equator being
$\frac{a}{2}=\frac{\pi}{\sqrt{n}}$, thus together
$\frac{3\pi}{2\sqrt{n}}$.} \label{poludnik}
\end{figure}

 We will show that for any probability distribution on the $n$
aforementioned states with min-entropy at least
$\log_2(n)-c$, the adversary's probability to determine the
plaintext is at most $p_{opt}+\epsilon$, where $\epsilon$
drops as $n^{-1/2}$. Let us
fix a particular distribution $(q_{i})_{i=1}^{n}$ on states ${|\psi_{i}%
\rangle}$. We construct a continuous distribution $\mu_{q}(|\phi\rangle)=\frac{nq_{i}%
}{4\pi}$ for $\phi\in{\mathcal{H}}_{i}$. The adversary's
probability to obtain the plaintext in the case of distribution
$(q_{i})_{i=1}^{n}$ on states ${|\psi_{i}\rangle}$ reads (compare
to Eq. \eqref{p_opt})
\begin{equation}
q_{n}=\sum_{i=1}^n q_{i}\left\vert \langle0|\psi_{i}%
\rangle\right\vert ^{2}.\label{qn}
\end{equation}
As the difference of the fidelity of the state ${|\psi_{i}%
\rangle}$ with ${|0%
\rangle}$ and any state from its neighborhood with ${|0%
\rangle}$ is upper bounded by $O\left( n^{-1/2}\right)$, we can
approximate Eq. (\ref{qn}) by a correctly normalized integral over
its neighborhood
\begin{equation}
\left\vert \langle0|\psi_{i}%
\rangle\right\vert ^{2}\leq
\frac{n}{4\pi}\int_{{|\phi\rangle}\in{\mathcal{H}}_{i}}\left[
\left\vert \langle0|\phi\rangle\right\vert ^{2}+O\left(
n^{-1/2}\right) \right]d\phi.
\end{equation}
Substituting we get%
\begin{align}
q_{n} &  \leq\sum_{i=1}^{n}q_{i}
\frac{n}{4\pi}\int_{{|\phi\rangle}\in{\mathcal{H}}_{i}}\left[
\left\vert \langle0|\phi\rangle\right\vert ^{2}+O\left(
n^{-1/2}\right) \right]d\phi\nonumber\\
&  =\sum_{i=1}^{n}
\int_{{|\phi\rangle}\in{\mathcal{H}}_{i}}\left[
q_{i}\frac{n}{4\pi}\left\vert \langle0|\phi\rangle\right\vert ^{2}+
q_{i}\frac{n}{4\pi}O\left(n^{-1/2}\right) \right]d\phi\nonumber\\
 &  =
\int_{{|\phi\rangle}\in{\mathcal{H}}}
\mu_q(|\phi\rangle)\left\vert \langle0|\phi\rangle\right\vert ^{2}d\phi +
\sum_{i=1}^{n}\left(q_i\frac{n}{4\pi}O\left(n^{-1/2}\right)
\int_{{|\phi\rangle}\in{\mathcal{H}}_{i}} d\phi\right)\nonumber\\
 &  \leq
p_{opt} +
\sum_{i=1}^{n}\left(q_iO\left(n^{-1/2}\right)\right)\label{mu}\\
 &  =
p_{opt} +
O\left(n^{-1/2}\right)\nonumber\\
 &  \leq\left[  1+O\left(  n^{-1/2}\right)  \right]  p_{opt}.\label{discretization}%
\end{align}
Inequality \eqref{mu} holds, because $\mu_{q}$ satisfies all
min-entropy requirements set in the previous section, and thus the integral can be bounded from above by $p_{opt}$.
The last inequality holds, because $p_{opt}$ is a positive constant, not depending on $n$.

% Note that the distribution $\mu_{q}$ satisfies all
%min-entropy requirements set in the previous section, so the
%adversary's probability to determine the plaintext is bounded by
%\[
%q_{n}\leq\left[  1+O\left(  n^{-1/2}\right)  \right]  \int_{{|\phi\rangle}%
%\in{\mathcal{H}}}\mu_{q}({|\phi\rangle})\left\vert \langle0|\phi
%\rangle\right\vert ^{2}d\phi\leq\left[  1+O\left(  n^{-1/2}\right)
%\right] p_{opt}.
%\]

We can conclude that for any $c\geq0$ and an arbitrarily small
$\epsilon>0$ there exist a (sufficiently large) $n$ that bounds
the adversary's probability to determine the original plaintext by
$1-2^{-c-1}+\epsilon$. It is obvious that for any fixed $c$ a
suitable $\epsilon$ can be chosen such that the probability of the
adversary to guess the ciphertext correctly is strictly smaller
than one.

 Moreover, consider a cryptosystem with only two elements $k_{1},k_{2}\in K$ such that
$Pr({\mathbf{K}}=k_{1})>0$, $Pr({\mathbf{K}}=k_{2})>0$. Then the
average states $\rho_{0}$, $\rho_{1}$ are not pure, i.e. (in a
suitable basis)
\begin{equation}
\rho_{0}=%
\begin{pmatrix}
\ a\  & 0\\
0 & 1-a
\end{pmatrix}
\end{equation}
with $a,1-a>0$. This implies the quantity in Eq.
(\ref{adversary_probability_quantum}) is strictly smaller
than $1$. Thus, for any encoding the adversary's
probability to determine the original plaintext is strictly
smaller than $1$, as long as there are at least two keys
with nonzero probability\footnote{Equivalently: Shannon (or
min-entropy) of the key is non-zero.}.
%%%%%%%%%%%%%%%%%%%%%%%%%%%%%%%%%%%%%%%%%%%%%%%%%%%%%%%%%%%%%%%%%%%%%%%%%%%%%%%

\section{Conclusion}

%%%%%%%%%%%%%%%%%%%%%%%%%%%%%%%%%%%%%%%%%%%%%%%%%%%%%%%%%%%%%%%%%%%%%%%%%%%%%%%
In this paper we have shown that a quantum ciphertext can
significantly increase the security of classical
communication between two parties using weak random
sources. In particular, for significantly weak sources,
where less than a quarter of keys can be used for
encryption, no security can be achieved with a classical
ciphertext. This is true independently both on the length
of the key $l$ and the length of the ciphertext. On the
contrary, for quantum ciphertexts, some level of security can
be achieved even if only two keys appear with non-zero probability.

For all sources (for all values of $c$) the presented
quantum approach is at least as good as the classical one.
Moreover, except for $c=0$ (trivial case with perfect
sources and perfect security) and $c=1$ the quantum
approach outperforms the classical for all values of $c$.

It is important to stress that our result does not give a lower bound on the
security one can achieve in this problem in general. Using encoding into
systems of higher dimension (e.g. a pair of qubits or a qutrit) may achieve
significantly better results, as more complicated encodings also using mixed
states come into question.

%%%%%%%%%%%%%%%%%%%%%%%%%%%%%%%%%%%%%%%%%%%%%%%%%%%%%%%%%%%%%%%%%%%%%%%%%%%%%%%

\section{Acknowledgments}

%%%%%%%%%%%%%%%%%%%%%%%%%%%%%%%%%%%%%%%%%%%%%%%%%%%%%%%%%%%%%%%%%%%%%%%%%%%%%%%

We acknowledge the support of the Czech Science Foundation GA{\v
C}R projects P202/12/1142 and P202/12/G061, as well as projects CE
SAS QUTE and VEGA 2/0072/12. MPl acknowledges the support of
SoMoPro project funded under FP7 (People) Grant Agreement no
229603 and by South Moravian Region.
%%%%%%%%%%%%%%%%%%%%%%%%%%%%%%%%%%%%%%%%%%%%%%%%%%%%%%%%%%%%%%%%%%%%%%%%%%%%%%%
%%%%%%%%%%%%%%%%%%%%%%%%%%%%%%%%%%%%%%%%%%%%%%%%%%%%%%%%%%%%%%%%%%%%%%%%%%%%%%%

 \nonumsection{References}
%GATHER{../../../../bibliografie/qcrypto.bib}
%GATHER{../../../../bibliografie/crypto.bib}
%GATHER{../../../../bibliografie/mathematics.bib}
%\bibliographystyle{splncs}
\bibliographystyle{plain}

\begin{thebibliography}{1}

\bibitem{BennetBrassard-1984}
Bennett, C. H. and Brassard, G.
\newblock Quantum cryptography: Public-key distribution and coin tossing,
\newblock {\em Proceedings of IEEE International Conference on Computers, Systems and Signal Processing}, pp. 175 - 179, 1984.

\bibitem{Blum-Independentunbiasedcoin-1984}
Manuel Blum.
\newblock Independent unbiased coin flips from a correlated biased source: A
  finite state markov chain.
\newblock In {\em Proceedings of the 25th Annual IEEE Symposium on Foundations
  of Computer Science}, pages 425--433, 1984.

\bibitem{ColbeckRenner-2011}
Roger Colbeck and Renato Renner.
\newblock No extension of quantum theory can have improved predictive power
\newblock arXiv:1005.5173v3 [quant-ph]

\bibitem{Dodis-2002}
Yevgeniy Dodis, and  Joel Spencer.
\newblock On the (non)Universality of the One-Time Pad.
\newblock In {\em Proceedings of the FOCS 2002}, pages 376--376, 2002.


\bibitem{Heinosaari-2009}
Teiko Heinosaari, Mario Ziman.
\newblock Guide to mathematical concepts of
quantum theory.
\newblock {\em Acta Physica Slovaca} 58, 487-674, 2009.

\bibitem{Helstrom-1976}
C. W. Helstrom.
\newblock Quantum
Detection and Estimation Theory.
\newblock {\em Academic Press} New York, 1976.


\bibitem{ChorGoldreich-Unbiasedbitsfrom-1988}
Benny Chor and Oded Goldreich.
\newblock Unbiased bits from sources of weak randomness and probabilistic
  communication complexity.
\newblock {\em SIAM Journal on Computing}, 17(2), pages 230--261, 1988.

\bibitem{idQuantique1}
http://www.idquantique.com/true-random-number-generator/products-overview.html, 25.5.2011

\bibitem{idQuantique2}
http://www.idquantique.com/network-encryption/cerberis-layer2-encryption-and-qkd.html, 25.5.2011

\bibitem{Kraetzel-LatticePoints-1989}
Ekkehard Krätzel.
\newblock {\em Lattice Points}.
\newblock Springer, 1989.

\bibitem{Lydersen-2010}
L. Lydersen, C. Wiechers, C. Wittmann, D. Elser, J. Skaar, and V. Makarov.
\newblock Hacking commercial quantum cryptography systems by tailored bright illumination.
\newblock {\em Nat. Photonics} 4, 686, 2010.

\bibitem{Mayers-2001}
D. Mayers
\newblock Unconditional security in quantum cryptography
\newblock {\em Journal of ACM}, vol 48 , 2001

\bibitem{McInnesPinkas-ImpossibilityofPrivate-1991}
J.~L. McInnes and B.~Pinkas.
\newblock On the impossibility of private key cryptography with weakly random
  keys.
\newblock In {\em Crypto'90}, pages 421--435, 1991.
\newblock LNCS 537.

\bibitem{Nielsen+Chuang-Quant_Compu_Quant:2000}
M.~A. Nielsen and I.~L. Chuang.
\newblock {\em {Quantum Computation and Quantum Information}}.
\newblock Cambridge University Press, Cambridge, 2000.

\bibitem{SanthaVazirani-Generatingquasi-randomsequences-1985}
Miklos Santha and Umesh~V. Vazirani.
\newblock Generating quasi-random sequences from slightly-random sources.
\newblock In {\em Proceedings of the 17th ACM Symposium on the Theory of
  Computing}, pages 366--378, 1985.

\bibitem{Shaltiel-RecentDevelopmentsin-2002}
Ronen Shaltiel.
\newblock Recent developments in explicit constructions of extractors.
\newblock {\em Bulletin of the EATCS}, 77, pages 67--95, 2002.

\bibitem{ShorPreskill-2000}
P. Shor and J. Preskill,
\newblock Simple Proof of Security of the BB84 Quantum Key Distribution Protocol
\newblock {\em Phys. Rev. Lett. 85}: 441-444, 2000

\bibitem{TamakiKoashiImoto-2003}
K. Tamaki, M. Koashi, and N. Imoto
\newblock Unconditionally Secure Key Distribution Based on Two Nonorthogonal States
\newblock {\em Phys. Rev. Lett. 90}, 2003

\bibitem{Neumann-Varioustechniquesused-1951}
John von Neumann.
\newblock Various techniques used in connection with random digits.
\newblock {\em National bureau of standards applied mathematics series},
  pages 36--38, 1951.

\end{thebibliography}

\end{document}